\documentclass[]{article}
\usepackage{spconf,amsmath,graphicx}
\usepackage{booktabs}
\usepackage{tabularx}
\usepackage{color}
\usepackage[colorlinks,bookmarks=true,citecolor=black,linkcolor=black,urlcolor=blue]{hyperref}

\title{Speech Emotion Recognition with Global-Aware Fusion on Multi-scale Feature Representation}
\name{Wenjing Zhu, Xiang Li\sthanks{Corresponding author.}}
\address{Du Xiaoman, Beijing, China \\ \{zhuwenjing02,lixiang01\}@duxiaoman.com}
\begin{document}
%
\maketitle
\begin{abstract}
Speech Emotion Recognition~(SER) is a fundamental task to predict the emotion label from speech data. 
Recent works mostly focus on using convolutional neural networks~(CNNs) to learn local attention map on fixed-scale feature representation by viewing time-varied spectral features as images.
However, rich emotional feature at different scales and important global information are not able to be well captured due to the limits of existing CNNs for SER. 
In this paper, we propose a novel \textbf{GL}obal-\textbf{A}ware \textbf{M}ulti-scale~(GLAM) neural network\footnote{The code is available at https://github.com/lixiangucas01/GLAM} to learn multi-scale feature representation with global-aware fusion module to attend emotional information. 
Specifically, GLAM iteratively utilizes multiple convolutional kernels with different scales to learn multiple feature representation.
Then, instead of using attention-based methods, a simple but effective global-aware fusion module is applied to grab most important emotional information globally.
Experiments on the benchmark corpus IEMOCAP over four emotions demonstrates the superiority of our proposed model with 2.5\% to 4.5\% improvements on four common metrics compared to previous state-of-the-art  approaches. 

\end{abstract}
\section{Introduction}\label{sec:intro}

Speech is the key medium of communication between people and plays an important role in everyone's daily life. Speech Emotion Recognition~(SER) aims to predict the emotion reflected in speech. SER is a fundamental task for various intelligent applications. For example, SER brings better user experience for intelligent robots by better understanding user's intents and states. Thus, interests from research communities have been increased for SER task~\cite{ELAYADI2011572}.

Recently, inspired by the uplifting progress in computer vision, existing studies \cite{CNNLSTM_Satt2017,DCNN2017,MCNN,DSCNN,Attnpooling,attnCNN_Zhang,headfusion2020,Area2020} have achieved great improvements on SER by viewing spectral features as images. The standard convolutional neural network~(CNN) architecture mainly consists of feature representation and attention map. Generally, feature representation usually learns fixed-scale features for capturing all informative features. Besides, attention map is utilized to attend most emotional information among local fixed-scale features \cite{Attnpooling,attnCNN_Zhang,headfusion2020,Area2020}. However, emotions are characterized by articulation variations with different scales in time-varied spectral feature.   

Due to the limits of existing CNNs in SER, we propose a novel \textbf{GL}obal-\textbf{A}ware \textbf{M}ulti-scale~(GLAM)  neural network. In our work, we utilize multi-scale convolutional blocks to extract features at varying scales due to different temporal spans and intonation strength in emotions. Furthermore, a global-aware fusion module is introduced to select important information reflecting emotional patterns in different scales. Experiments show both multi-scale feature representation and global-aware fusion can bring great improvements compared to the state-of-the-art CNN model\cite{Area2020}.

Our contributions can be summarized as follows:
\begin{itemize}
	\item To the best of our knowledge, we are the first to apply multi-scale feature representation to SER. Besides, this representation also enhances existing CNNs greatly. 
	\item We introduce a new global-aware fusion module to grab emotional information across multiple scales. Our study displays the great potential of information fusion among different granularities. 
	\item Experiments demonstrate the superiority of our proposed model on four common metrics for SER. We achieve 2.5\% to 4.5\% improvements on the IEMOCAP corpus compared to state-of-the-art methods. 
\end{itemize}

The rest of this paper is organized as follows. We review related methods in Section~\ref{sect_relatedwork}. Section~\ref{sect_model} presents the proposed GLAM neural network. Finally, Section~\ref{sect_experiments} and Section~\ref{sect_conclusion} are the experiment results and conclusion.

\section{Related Work}\label{sect_relatedwork}

Typical works on SER focus on feature representation of the emotion information in speech. The first deep learning method was utilized to learn feature representation for predicting emotion label of each feature segment~\cite{han2014speech}. Then, by viewing feature segments as images, convolutional networks show the great effects on feature representation for SER~\cite{CNNLSTM_Satt2017,MCNN,Attnpooling,headfusion2020,Area2020,Chen2014CNNLSTM,Wu2019Capsule}. Finally, various methods such as extreme learning machine~(ELM) \cite{han2014speech}, capsule networks \cite{Wu2019Capsule}, and LSTM \cite{CNNLSTM_Satt2017,Chen2014CNNLSTM} have been applied to fuse all feature representations from all feature segments. However, most existing CNNs focus on fixed-scale local feature and ignore the effects of varied scales, which are fully considered in our proposed model.

Recently, attention-based models have made significant progress on SER~\cite{Attnpooling,attnCNN_Zhang,headfusion2020,Area2020,selfAttention2019}. 
For example, multi-head attention maps are learned by convolutional operations to select important information according to surrounding information~\cite{headfusion2020}. Moreover, area attention is further introduced to compute the importance from different ranges of convolutions~\cite{Area2020}. Though local information is useful for describing subtle information, the lack of perceiving global information limits the advance of existing attention-based methods. Hence, we propose a global-aware fusion module to compute global attention map and fuse emotional information across different scales.

\section{Methodology}\label{sect_model}
\begin{figure}[tb!]
  \centering
  \includegraphics[width=\columnwidth]{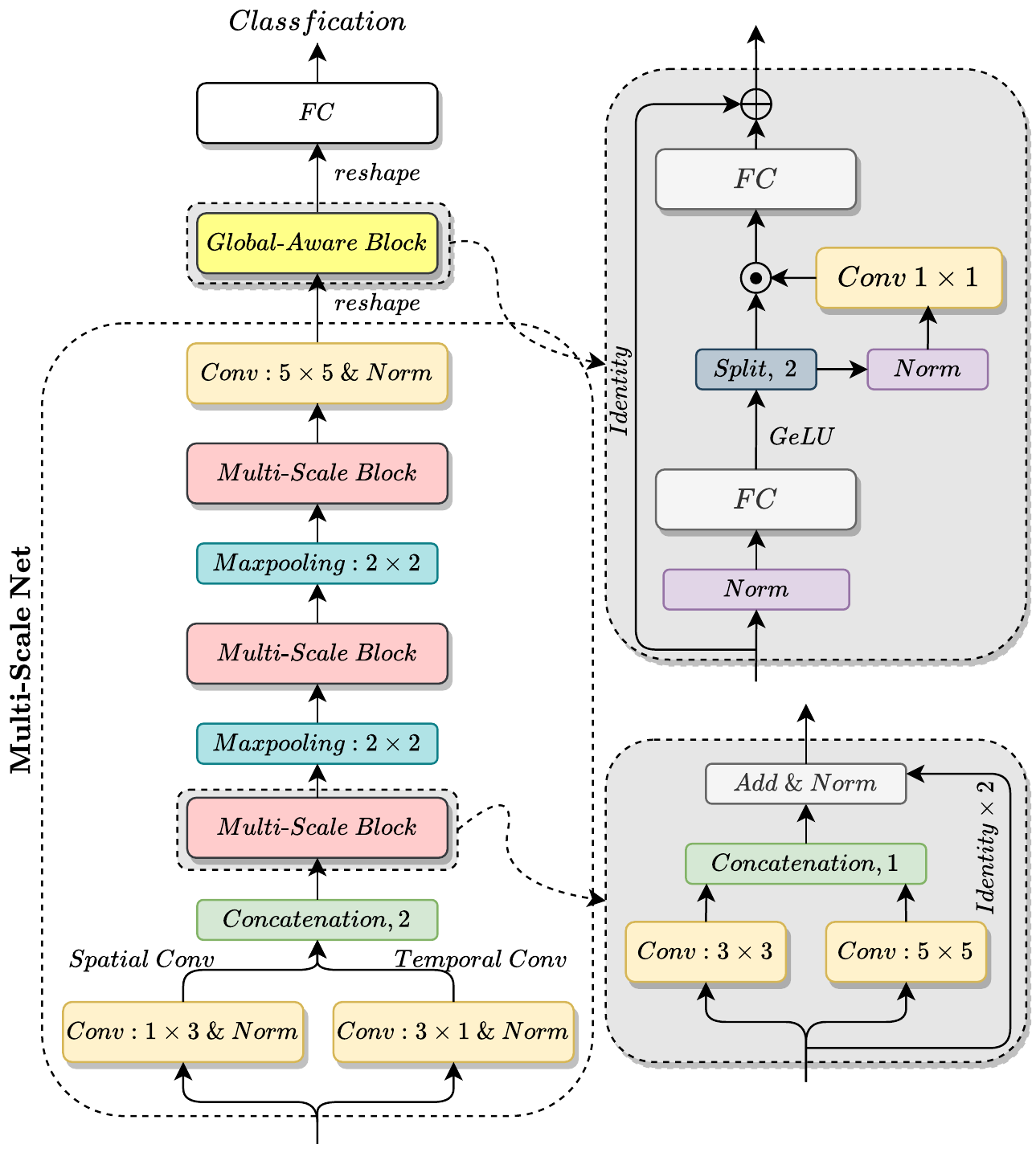}
  \caption{Overview of the model architecture.\label{fig:model}}
  \end{figure}

In this section, we present our GLAM model in detail. As shown in Fig.~\ref{fig:model}, GLAM contains two main modules: multi-scale module and global-aware fusion module. We first introduce these two modules and then combine a \textit{mixup} method to enhance the generalization of GLAM.

\subsection{Multi-scale Feature Representation}
 We propose a multi-scale block to parallelize multi-scale convolutional layers on the same level. The multi-scale block, which receive different sized receptive fields, can capture multi-scale feature representation.
The structure is shown in right down panel of Fig.~\ref{fig:model}. We keep the channel size unchanged. 
For better capturing spatial and temporal features, we implement kernel size of \(1\times 3\) for spatial convolution and kernel size of \(3\times 1\) for temporal convolution with batch normalization put after on. The out-channels both are 16. 
For the first layer, we concatenate spatial and temporal convolved outputs alongside spatial dimension. 
For the rest, the outputs are concatenated alongside channel dimension. 
In the last convolutional layer with batch normalization followed, we use a large kernel size of \(5\times5\). 
Same paddings are applied to all of convolutional layers. 
  
\subsection{Global-Aware Fusion}
As the multi-scale architecture extracts local features,  
we consider correlating features and enhance feature communications globally. 
We implement a gMLP block \cite{gMLP}, named as a global-aware block in this paper, shown in the right up panel of Fig. \ref{fig:model}, to enhance cross-scale feature communication. 
For a feature map X with a size of \(C \times d_f\), where C is the channel size and \(d_f\) represents size of flattened feature map. 
The output dimension of first and last fully connected layer are \(4d_f\) and \(d_f\) respectively. 
After a projection of GeLU, the output is split along feature dimension. 
And the convolutional layer keeps channel size unchanged. 
The operations of multiplication enhance feature mixing across channel dimension. 
The output of global-aware fusion module is then reshaped to feed into fully connected (FC) network for classification. 

\subsection{Data Augment}
To improve generalization capability of the system, we use a \textit{mixup} method \cite{zhang2018mixup} for training, which combines pairs of examples and their labels. 
\textit{Mixup} mixes two pair training examples \((x_i, y_i)\) and \((x_j, y_j)\) as a new constructed example
\begin{eqnarray}
  \tilde{x} = \lambda x_i + ( 1 - \lambda) x_j \\
  \tilde{y} = \lambda y_i + ( 1 - \lambda ) y_j
\end{eqnarray}
where \((x_i, y_i)\) and \((x_j, y_j)\) are randomly drawn from training data 
and \( \lambda \sim Beta(\alpha, \alpha) \) with \( \alpha \in (0, \infty) \). 
\textit{Mixup} regularizes the neural network to favor simple linear behavior in-between training examples. 
This method effectively smoothens discrete data space into continuous space. 

\begin{table*}[hbtp]
  \caption{\label{tab:coma} Comparison of evaluation metrics on three types of datasets.}
  \centering
  \begin{tabular}{cccccc}
    \toprule
    Dataset & Model & WA & UA & macro F1 & micro F1 \\
    \midrule
    & APCNN & 70.73\(\pm\)2.57 & 67.96\(\pm\)2.66 & 68.26\(\pm\)3.17 & 70.20\(\pm\)3.00 \\
    Improvisation
    & MHCNN & 76.09\(\pm\)2.01 & 73.87\(\pm\)2.31 & 74.27\(\pm\)2.25 & 75.91\(\pm\)2.05 \\
    & AACNN & 78.47\(\pm\)2.42 & 76.68\(\pm\)3.29 & 76.69\(\pm\)3.27 & 78.29\(\pm\)2.58 \\
    & \textbf{GLAM} & \textbf{81.18\(\pm\)1.47} & \textbf{79.25\(\pm\)1.88} & \textbf{79.87\(\pm\)1.64} & \textbf{80.99\(\pm\)1.50} \\
    \midrule
    & APCNN & 55.95\(\pm\)2.93 & 54.86\(\pm\)2.74 & 48.75\(\pm\)3.71 & 51.15\(\pm\)3.87 \\
    Script
    & MHCNN & 64.57\(\pm\)2.12 & 63.42\(\pm\)2.14 & 61.46\(\pm\)2.50 & 63.36\(\pm\)2.56 \\
    & AACNN & 67.20\(\pm\)4.78 & 65.94\(\pm\)6.19 & 64.82\(\pm\)7.63 & 66.55\(\pm\)6.99 \\
    & \textbf{GLAM} & \textbf{71.44\(\pm\)2.05} & \textbf{70.39\(\pm\)2.13} & \textbf{69.56\(\pm\)2.13} & \textbf{70.91\(\pm\)2.12} \\
    \midrule
    & APCNN & 62.53\(\pm\)1.43 & 62.80\(\pm\)1.53 & 61.89\(\pm\)1.52 & 62.12\(\pm\)1.50 \\
    Full
    & MHCNN & 69.80\(\pm\)1.69 & 70.09\(\pm\)1.74 & 69.65\(\pm\)1.73 & 69.68\(\pm\)1.71 \\
    & AACNN & 70.94\(\pm\)5.08 & 71.04\(\pm\)4.84 & 70.59\(\pm\)6.34 & 70.71\(\pm\)6.44 \\
    & \textbf{GLAM} & \textbf{73.70\(\pm\)1.25} & \textbf{73.90\(\pm\)1.31} & \textbf{73.51\(\pm\)1.29} & \textbf{73.60\(\pm\)1.27} \\
    \bottomrule
  \end{tabular}
\end{table*}

\begin{table*}[t]
  \caption{\label{tab:Ablation} Performance of multi-scale module and global-aware module on Improvisation dataset.}
  \centering
  \begin{tabular}{ccccc}
   \toprule
   Model & WA & UA & macro-F1 & weighted F1 \\
   \midrule
  \textbf{GLAM} & \textbf{81.18\(\pm\)1.47} & \textbf{79.25\(\pm\)1.88} & \textbf{79.87\(\pm\)1.64} & \textbf{80.99\(\pm\)1.50} \\
  Multi-scale & 80.89\(\pm\)1.43 & 78.85\(\pm\)1.65 & 79.47\(\pm\)1.60 & 80.67\(\pm\)1.45 \\
  Multi-scale + MHA & 80.43\(\pm\)1.53 & 78.92\(\pm\)1.92 & 79.20\(\pm\)1.68 & 80.26\(\pm\)1.53 \\
  Multi-scale + AA & 80.61\(\pm\)1.75 & 79.09\(\pm\)1.75 & 79.47\(\pm\)1.80 & 80.46\(\pm\)1.77 \\
  MHCNN & 76.09\(\pm\)2.01 & 73.87\(\pm\)2.31 & 74.27\(\pm\)2.25 & 75.91\(\pm\)2.05 \\
  AACNN & 78.47\(\pm\)2.42 & 76.68\(\pm\)3.29 & 76.69\(\pm\)3.27 & 78.29\(\pm\)2.58 \\

  \bottomrule
  \end{tabular}
\end{table*}

\begin{table*}[bt!]
  \caption{\label{tab:alpha}Result of \(\alpha\) on Improvisation dataset.}
  \centering
  \begin{tabular}{ccccc}
   \toprule
   \(\alpha\) & WA & UA & macro-F1 & weighted F1 \\
   \midrule
  0  & 80.44\(\pm\)1.55 & 78.86\(\pm\)1.65 & 79.32\(\pm\)1.56 & 80.32\(\pm\)1.56 \\
  0.3 & 81.15\(\pm\)1.73 & 79.26\(\pm\)1.98 & 79.88\(\pm\)1.85 & 80.98\(\pm\)1.76 \\
  \textbf{0.5} & \textbf{81.18\(\pm\)1.47} & \textbf{79.25\(\pm\)1.88} & \textbf{79.87\(\pm\)1.64} & \textbf{80.99\(\pm\)1.50} \\
  0.8 & 80.97\(\pm\)1.53 & 78.97\(\pm\)1.85 & 79.62\(\pm\)1.75 & 80.79\(\pm\)1.53 \\
  1 & 80.96\(\pm\)1.72 & 79.19\(\pm\)1.92 & 79.65\(\pm\)1.88 & 80.78\(\pm\)1.73 \\
  2 & 80.80\(\pm\)1.39 & 79.18\(\pm\)1.66 & 79.55\(\pm\)1.50 & 80.64\(\pm\)1.40 \\
  3 & 80.61\(\pm\)1.45 & 78.95\(\pm\)1.58 & 79.26\(\pm\)1.59 & 80.44\(\pm\)1.45 \\
   \bottomrule
  \end{tabular}
\end{table*}

\section{Experiments} \label{sect_experiments}
In this section, we first introduce the experimental settings. Then, we compare our proposed model with state-of-the-art methods. Finally, we conduct ablation study on two modules.

\subsection{Experimental Settings}

\textbf{Datasets.} We conduct our experiments on the benchmark corpus IEMOCAP~\cite{busso2008iemocap}. IEMOCAP contains 12 hours of emotional speech performed by 10 actors. We select four types of emotions \textit{angry}, \textit{happy}, \textit{sad}, and \textit{neutral} as same as previous studies~\cite{Attnpooling,attnCNN_Zhang,headfusion2020,Area2020,selfAttention2019} for balancing the category distribution. 
In agreement with previous works, we label \textit{excited} as \textit{happy} category due to similarity of two categories. 
According to whether the actors perform a fixed script, we have three types of datasets for experiments: Improvisation, Script, and Full. 

\noindent\textbf{Compared Methods and Evaluation Metrics.} To study our proposed model deeply, we choose the best model of Area Attention CNN~(AACNN) \cite{Area2020} with recent work Multi-Head Attention CNN~(MHCNN) \cite{headfusion2020} and Attention Pooling CNN~(APCNN) \cite{Attnpooling} for comparison. Moreover, to comprehensively understand prediction performance, we utilize four common metrics Weighted Accuracy~(WA), Unweighted Accuracy~(UA), Micro F1 score, and Macro F1 score. The distinction between WA and UA~(
micro and macro F1 score) is that the former takes label imbalance into account, while the latter does not. We report the mean value and standard deviation of above metrics in experiments. In addition, we will also show confusion matrix for concrete analysis. 

\noindent\textbf{Implementation Details.} Since there is no unified way to split dataset, we randomly divide dataset into 80\% of data for training and the rest for test with 100 times to ensure the reliability of experiment results. 
We use mel-frequency cepstral coefficients (MFCCs) \cite{Ayadi2007ICASSP,DAI2015777} as the feature input.
Each utterance is divided into 2-second segments with 1.6 second overlap between segments. We average the prediction results of all segments in same utterance as the prediction result. 
Cross-entropy criterion is used as an objective function. Adam optimizer is adopted with weight decay rate as \(10^{-6}\). 
The learning rate is initialized to \(10^{-4}\) and exponentially decayed with multiplicative factor 0.95 until the value reaches $10^{-6}$. The models are all trained for 50 epochs and the batch size is set to 32. In our proposed model, we evaluate the performance of different mixup parameter $\alpha$ and set it to $0.5$ for comparison. 

\begin{figure}[bthp]
  \centering
  \includegraphics[width=1\columnwidth]{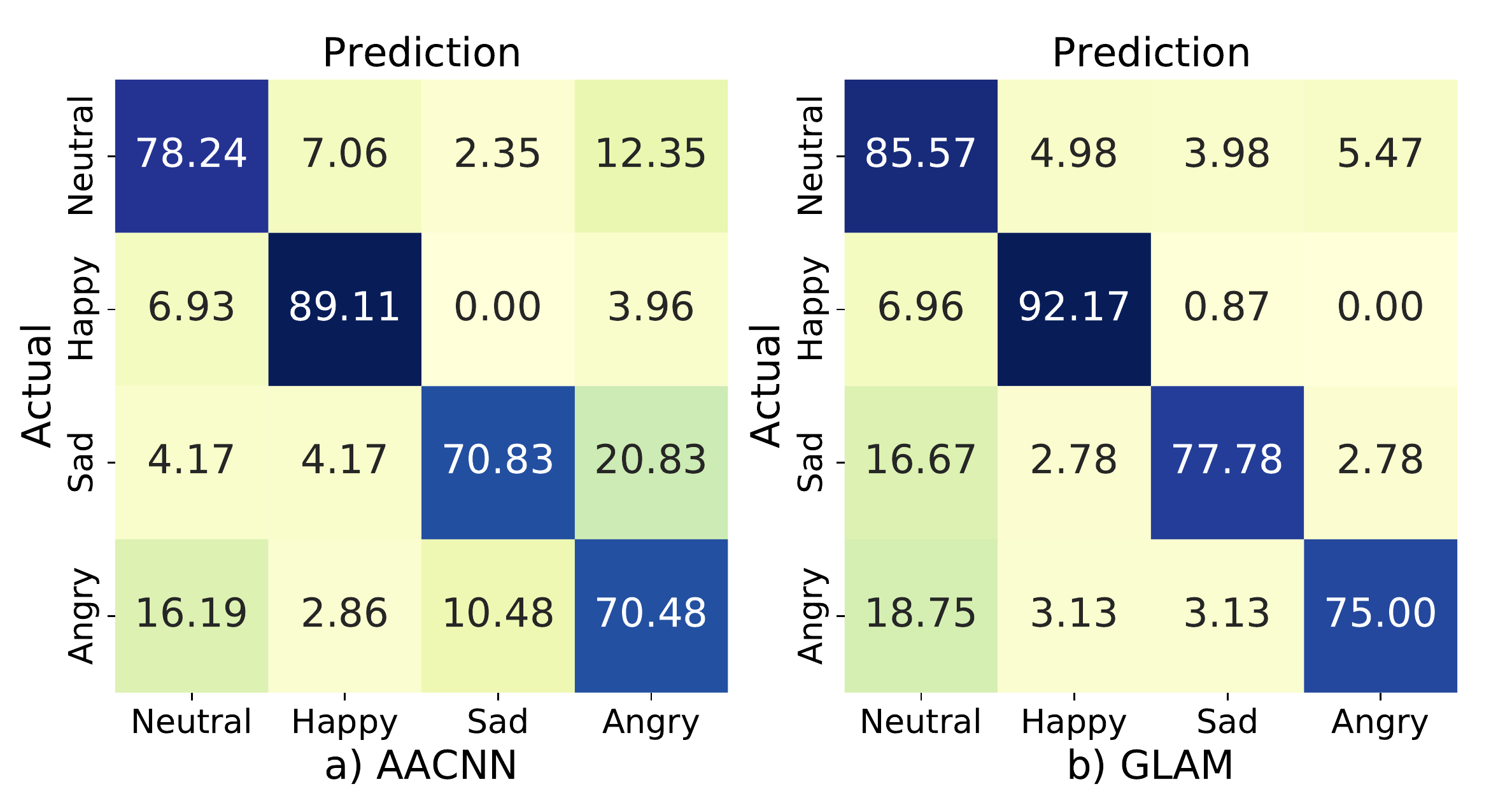}
  \caption{Confusion matrix on Improvisation dataset. a) and b) represent results of AACNN and GLAM model respectively.  \label{fig:confusion}}
\end{figure}

\subsection{Results and Discussion}
Experiments are conducted on three types of datasets with four common metrics, shown in Table~\ref{tab:coma}. 
Compared to recent the state-of-the-art approaches, the proposed architecture significantly improves performances on all listed metrics, which demonstrates the effectiveness of global-aware fusion and multi-scale technique. 

\begin{figure}[htb]
    \centering
    \includegraphics[width=1\columnwidth]{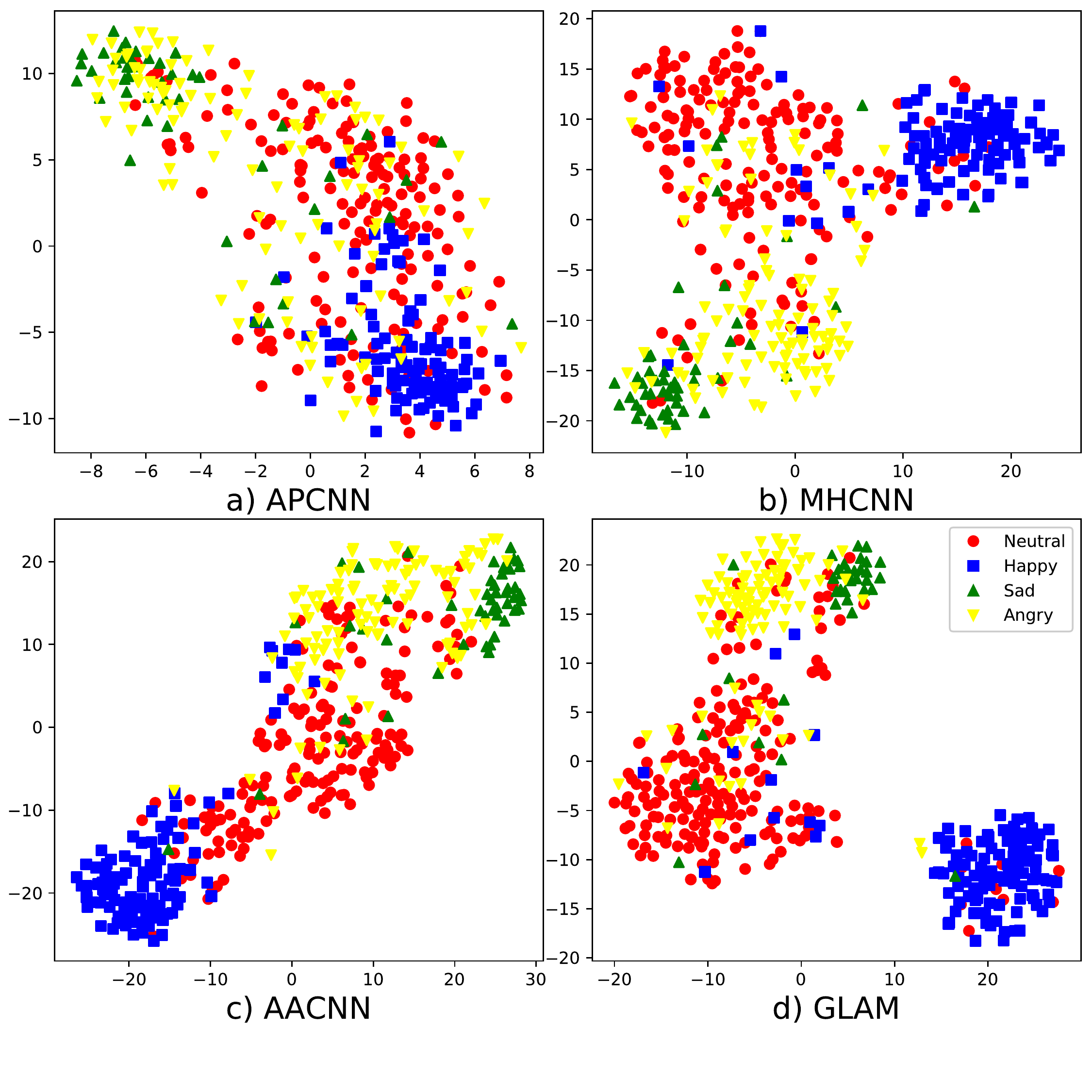}
    \caption{Visualizations of hidden states by commonly-used t-SNE method\cite{tsne} on improvisation dataset for four models: a) APCNN; b) MHCNN; c) AACNN; d) GLAM.\label{fig:tsne}}
    \vspace{-0.5cm}
\end{figure}

Meanwhile, GLAM has smaller values of standard deviation for four common metrics which indicates a robust and stable performance. 
In detail, we compute a confusion matrix to exploit accuracies for each emotional category, see in Fig.~\ref{fig:confusion}. 
GLAM eliminates confusion sets of \textit{neutral-angry} and \textit{sad-angry} and superiorly increase 6.95\% on prediction of \textit{sad} label and 4.52\% on prediction of \textit{angry} label. 
With t-distributed stochastic neighbor embeddings (t-SNE) method \cite{tsne}, we show 2D projections of high-level features in Fig.~\ref{fig:tsne} by using the last second layer output for APCNN, MHCNN and GLAM and applying an additional dense layer to resize large sized output for AACNN. 
The visualization of APCNN in Fig. \ref{fig:tsne} a) shows a drastic overlap among four categories. 
For a MHCNN model shown in Fig. \ref{fig:tsne} b), the \textit{happy} category is unobviously departed from other categories. 
For a AACNN model shown in Fig. \ref{fig:tsne} c), the areas of four categories are successively joint together. 
In our model, see in Fig. \ref{fig:tsne} d), the \textit{happy} category is apparently separated from other categories. We see a few \textit{angry} points located in area of \textit{neutral}. 
There is a better distinction between \textit{sad} and \textit{angry} categories compared to AACNN model. 

From above results, we see effectiveness of the multi-scale technique with a global-aware perceiving. 
Previous works only scale features with fixed length. 
And current attention-based SER methods are lack of perceiving global information. 
While, GLAM exploits features at different scales for multiple spanned patterns and attend global information across different features. 


\subsection{Ablation Study}
We conduct ablation experiments on the model by removing and replacing modules used in our proposed module. 
Results are shown in Table \ref{tab:Ablation}. 
By removing the global-aware fusion module, the multi-scale module universally outperforms over all existing methods. 
By replacing such module with Multi-Head Attention (MHA) \cite{headfusion2020} or Area Attention (AA) \cite{Area2020}, 
the performance even degrades compared to pure multi-scale module, which indicates effective attention on high dimensional representation globally rather than local fusion. 

\noindent\textbf{Selection of \(\alpha\).}
Without the \textit{mixup} method~(i.e. \(\alpha = 0\)), our model surpasses previous works. 
Further improvement is realized by using \textit{mixup} method with the optimal result for \(\alpha = 0.5\) in Table \ref{tab:alpha}, which indicates the effectiveness of enhancing generalization capability.

\section{Conclusion}\label{sect_conclusion}
In this paper, we propose a novel architecture, named as GLAM network, which combines a multi-scale module and a global-aware fusion module. 
The multi-scale module captures different scaled features for various time-spanned patterns and the global-aware module enhances cross-scale communication and attends feature confusion at different scales.
In addition, we use \text{mixup} method to enhance the generalization of our model. Experiment results demonstrate that our model achieves the state-of-the-art performance on the benchmark dataset IEMOCAP over 4-class emotion classification. 


\bibliographystyle{IEEEbib}
\bibliography{ser,attention,model}

\appendix
\section{Supplementary}
We stress that there is no unified and regular way to split dataset. And it is also unstable to for training if we set different random seeds. Furthermore, the high accuracy may be obtained by choosing one optimal partition way of dataset.
To ensure the reliability and strictness of experiment results, we randomly split dataset 100 times for easily reproducing
this work. 
We sincerely apologize for leaving out the validation set. Indeed, the validation set is truly necessary. Since recent works \cite{headfusion2020,Area2020} omitted it due to scarcity of dataset, we merely follow the split strategy for better comparison. 
According to reviewers' comments, also for the sake of strictness, we randomly divide the dataset into training, validation, and test sets at a ratio of 8:1:1 for comparison. 
During training, we save the optimal result according to the best mean value of UA and WA metrics on the validation set. 
Results are listed in the Table. \ref{tab:coma1}. 
We see our model also achieve the state-of-the-art performance with a robust stability.

\begin{table*}[hbtp]
  \caption{\label{tab:coma1} Comparison of evaluation metrics on three types of datasets with a validation set.}
  \centering
  \begin{tabular}{cccccc}
    \toprule
    Dataset & Model & WA & UA & macro F1 & micro F1 \\
    \midrule
    & APCNN & 69.93 \(\pm\) 4.33 & 62.70 \(\pm\) 5.16 & 63.04 \(\pm\) 5.94 & 68.86 \(\pm\) 5.48 \\
    Improvisation
    & MHCNN & 76.13 \(\pm\) 2.80 & 71.15 \(\pm\) 4.11 & 71.91 \(\pm\) 3.92 & 75.92 \(\pm\) 2.87 \\
    & AACNN &  78.65 \(\pm\)3.30 & 74.31 \(\pm\)4.59 & 74.78 \(\pm\)4.43 & 78.47 \(\pm\)3.40\\
    & \textbf{GLAM} & \textbf{81.04 \(\pm\) 2.65} & \textbf{75.89 \(\pm\) 3.94} & \textbf{76.79 \(\pm\) 3.78} & \textbf{80.78 \(\pm\) 2.75} \\
    \midrule
    & APCNN & 53.48 \(\pm\) 3.83 & 55.77 \(\pm\) 3.60 & 48.14 \(\pm\) 4.27 & 48.10 \(\pm\) 4.90 \\
    Script
    & MHCNN & 63.03 \(\pm\) 3.52 & 64.37 \(\pm\) 3.24 & 61.32 \(\pm\) 3.88 & 61.77 \(\pm\) 4.31 \\
    & AACNN &  65.49 \(\pm\)3.49 & 65.87 \(\pm\)2.98 & 64.36 \(\pm\)3.42 & 65.22 \(\pm\)3.77\\
    & \textbf{GLAM} & \textbf{69.80 \(\pm\) 3.16} & \textbf{70.63 \(\pm\) 3.17} & \textbf{68.85 \(\pm\) 3.40} & \textbf{69.30 \(\pm\) 3.42} \\
    \midrule
    & APCNN & 60.33 \(\pm\) 2.67 & 62.16 \(\pm\) 2.55 & 60.03 \(\pm\) 2.78 & 59.67 \(\pm\) 2.88 \\
    Full
    & MHCNN & 67.67 \(\pm\) 2.34 & 68.65 \(\pm\) 2.32 & 67.55 \(\pm\) 2.34 & 67.58 \(\pm\) 2.36 \\
    & AACNN &  67.71 \(\pm\)2.25 & 68.67 \(\pm\)2.22 & 67.62 \(\pm\)2.32 & 67.46 \(\pm\)2.33\\
    & \textbf{GLAM} & \textbf{71.63 \(\pm\) 2.12} & \textbf{72.56 \(\pm\) 2.05} & \textbf{71.53 \(\pm\) 2.16} & \textbf{71.42 \(\pm\) 2.19} \\
    \bottomrule
  \end{tabular}
\end{table*}

\end{document}